\documentclass[twocolumn,floatfix,pra,aps,showpacs]{revtex4}
\usepackage{epsfig,graphicx,tabularx}
\usepackage{subfigure}
\usepackage{amsmath,amsfonts,amssymb}
\usepackage{color,braket}

\newcommand\ii{{\rm i}}

\begin{document}

\title{Ground-state properties and elementary excitations of quantum droplets in dipolar Bose-Einstein condensates}
\author{F. W\"achtler}
\author{L. Santos}
\affiliation{Institut f\"ur Theoretische Physik, Leibniz Universit\"at Hannover, Appelstr. 2, DE-30167 Hannover, Germany}

\begin{abstract}
Recent experiments have revealed the formation of stable droplets in dipolar Bose-Einstein condensates. This surprising result has been explained 
by the stabilization given by quantum fluctuations. We study in detail the properties of a BEC in the presence of quantum stabilization.
The ground-state phase diagram presents three main regimes: mean-field regime, in which the quantum correction is perturbative, 
droplet regime, in which quantum stabilization is crucial, and a multi-stable regime. In the absence of a multi-stable region, the condensate undergoes 
a crossover from the mean-field to the droplet solution marked by a characteristic growth of the peak density that may be employed to clearly distinguish quantum 
stabilization from other stabilization mechanisms. Interestingly quantum stabilization allows for three-dimensionally self-bound condensates. We characterized these 
self-bound solutions, and discuss their realization in experiments. We conclude with a discussion of the lowest-lying excitations both for trapped condensates, and 
for self-bound solutions.
\end{abstract}


\maketitle

\section{Introduction}
\label{sec:Intro}

Ultra cold dipolar gases, in which magnetic or electric dipole-dipole interactions~(DDI) play a crucial role, differ substantially from their non-dipolar counterparts. 
The long-range anisotropic nature of the DDI, and the interplay between DDI and short-range interactions, 
leads to rich new physics~\cite{Lahaye2009,Baranov2012}, which is just 
starting to be unveiled in experiments on magnetic atoms~\cite{Griesmaier2005,Lu2011,Aikawa2012,DePaz2013}, 
polar molecules~\cite{Ni2008,Yan2013,Takehoshi2014,Park2015}, and Rydberg-dressed atoms~\cite{Rydberg-Dressed}. 
Since the DDI are partially attractive, dipolar gases may undergo instability unless the DDI are properly compensated by the contact interaction. 
Chromium experiments showed that, as in non-dipolar Bose-Einstein condensates~(BECs) with negative $s$-wave scattering length, $a<0$, 
an unstable dipolar BEC collapses when the scattering length is quenched below a critical value~\cite{Lahaye2008}.

Surprisingly, recent Dysprosium experiments~\cite{Kadau2015,Ferrier2016} have revealed that destabilization of a dipolar BEC does not generally lead to collapse, as previously assumed.
In these breakthrough experiments destabilization leads to the formation of stable droplets, that are only destroyed in a long time scale by three-body losses. 
Although first studies pointed to the possibility that large three-body conservative forces could stabilize the droplets~\cite{Xi2015,Bisset2015}, recent works 
have shown that the most plausible stabilization mechanism is due to quantum fluctuations~\cite{Ferrier2016,Wachtler2016,Saito2016}, which 
play a similar role as that of surface tension in classical ferrofluids~\cite{Cowley1967,Timonen2013}. 

The stabilization mechanism, that stems from the corresponding Lee-Huang-Yang~(LHY) correction of the condensate energy, 
results from the dipolar character of the condensate~\cite{Wachtler2016}. The anisotropic DDI  result in hard~(soft) modes depending on whether the momentum of the excitation is quasi-parallel~(perpendicular) 
to the dipole moment. Whereas long-wave length soft modes drive the BEC instability at sufficiently low $a$, hard modes dominate the repulsive LHY correction, resembling the 
situation recently studied in binary Bose-Bose mixtures~\cite{Petrov2015}. When the BEC becomes unstable, the local growth of the density $n$ associated to 
the modulational instability is eventually halted, since the repulsive LHY correction of the chemical potential scales as $n^{3/2}$ compared to the $n$ dependence of the mean-field term. 
Crucially, this occurs despite a very small condensate depletion. The latter is possible since in the vicinity of the instability the mean-field term is very small 
due to the quasi-compensation of the DDI and contact interaction. Quantum stabilization and the associated droplet nucleation are hence characteristic features of strongly dipolar BECs, which 
were absent in Chromium experiments only due to the relatively weak DDI~\cite{Wachtler2016}.

This paper analyzes the ground-state properties and excitations of dipolar BECs in the regime where LHY stabilization becomes relevant, both by means 
of numerical simulations of the corresponding generalized non-linear non-local Schr\"odinger equation~(gNLNLSE), and by a Gaussian ansatz approach.
We first analyze the ground-state as a function of $a$ and the condensate aspect ratio, that, as for the case of three-body stabilization~\cite{Blakie2016}, 
splits in three regimes: mean-field, droplet, and multi-stable, the latter occurring only for sufficiently pancake traps.  We focus then in the regime without multi-stability region, in which 
the mean-field solution undergoes for decreasing $a$ a cross-over into the droplet regime. This crossover is marked by a characteristic growth of the peak density whose functional form 
differs from that expected for three-body stabilization. The LHY stabilization results in a three-dimensional~(3D) self-bound~(SB) solution, which we characterize, discussing as well 
the conditions for its observability. Finally, we characterize the collective excitations of the dipolar BEC in the presence of LHY stabilization both for trapped BECs and in the SB regime.

The paper is structured as follows. Section~\ref{sec:Model} introduces both the gNLNLSE and the variational Gaussian ansatz. Section~\ref{sec:Ground} is devoted to the analysis of the ground-state properties, 
including the ground-state regimes, and the evolution of the peak density in the crossover regime between mean-field and droplet regimes.
In Sec.~\ref{sec:SB} we analyze the properties and conditions for observability of 3D SB condensates.
Section~\ref{sec:Excitations} discusses the lowest-lying excitations of both trapped BECs and SB condensates.
Finally Sec.~\ref{sec:Conclusions} summarizes our conclusions.



\section{Model}
\label{sec:Model}

\subsection{Generalized non-local non-linear Schr\"odinger equation}

We consider a harmonically trapped BEC of $N$ magnetic dipoles of mass m and dipole moment ${\boldsymbol \mu}$ oriented along the 
$z$ direction by an external magnetic field~(equivalent results can be found for electric dipoles). 
In order to describe dipolar BECs including the effect of quantum fluctuations, we recently introduced the gNLNLSE~\cite{Wachtler2016}:
\begin{eqnarray}
\ii\hbar \frac{\partial}{\partial t}\psi({\bf r},t) &=&  \left [ -\frac{\hbar^2}{2m}\nabla^2+V({\bf r})+gn({\bf r},t)   \right\delimiter 0 \nonumber \\ 
 && \left\delimiter 0  +\int d^3r'V_{\rm dd}({\bf r}-{\bf r}')n({\bf r}',t) \right\delimiter 0 \nonumber \\ 
 && \left \delimiter 0 + g_{\rm LHY}n({\bf r},t)^{3/2}  \right ]             \psi({\bf r},t), 
 \label{GPE}
\end{eqnarray}
where $\psi$ denotes the condensate wavefunction~(with $\int d^3r |\psi({\bf r},t)|^2=1$),  $V({\bf r})=\frac{m}{2}\left(\omega_x^2 x^2+\omega_y^2 y^2+\omega_z^2 z^2\right)$ is the trapping potential
with $\omega_{x,y,z}$ the trapping frequencies, and $n({\bf r},t)=|\psi({\bf r},t)|^2$. The contact interaction strength is given by $g=\frac{4\pi \hbar^2 N a}{m}$ where $a$ is the scattering length. The dipolar potential takes the form $V_{\rm dd}({\bf r})=N \frac{\mu_0 \mu^2}{4\pi}\frac{1-3\cos^2 \vartheta}{|{\bf r}|^3}$, where $\mu_0$ is the vacuum permeability and $\vartheta$ the angle between ${\bf r}$ and ${\boldsymbol \mu}$. 

Whereas the first two lines of Eq.~\eqref{GPE} correspond to the NLNLSE thoroughly employed for the study of dipolar condensates~\cite{Lahaye2009}, the last line stems from the LHY correction 
to the equation of state, which is obtained using the local density approximation~(LDA) from the knowledge of the LHY correction in homogeneous 3D space~ \cite{Lima2011,Lima2012}. 
The strength of the LHY correction is given by $g_{\rm LHY}=\frac{32}{3\sqrt{\pi}}g\sqrt{N a^3}F(\epsilon_{\rm dd})$, where $F(\epsilon_{\rm dd})=\frac{1}{2}\int d\vartheta_k \sin{\vartheta_k} f(\epsilon_{\rm dd},\vartheta_k)^{5/2}$ and $f(\epsilon_{\rm dd},\vartheta_k)=1+\epsilon_{dd}(3\cos^2\vartheta_k-1)$ with $\epsilon_{\rm dd}=N\frac{\mu_0 \mu^2}{3g}$. 

The validity of Eq.~\eqref{GPE}, and in particular of the LDA treatment of the LHY term, demands in principle that the system remains in the Thomas-Fermi~(TF) regime in all spatial directions. 
This is approximately the case for the relatively big droplets discussed in this paper. However, the equation remains valid even for smaller droplets, as long as the main contribution to the LHY term 
is provided by short-wavelength excitations~(see the discussion in Ref.~\cite{Wachtler2016}). In particular, recent numerical results based on Path Integral Monte Carlo techniques have confirmed the 
validity of Eq.~\eqref{GPE} for the description of the small droplets created in recent quench experiments~\cite{Saito2016}. 

Finally, we note that the use of Eq.~\eqref{GPE} is just valid for cigar-shape traps~(along $z$) or moderately pancake ones, in which (in absence of LHY stabilization) the 
BEC would be destabilized at $\epsilon_{dd}$ close to $1$~($\epsilon_{dd}=1$ marks the instability threshold for a 3D homogeneous dipolar BEC). 
For traps with larger aspect ratio, $\lambda=\omega_z/\omega_{x,y}$, the trap geometry significantly stabilizes the BEC. As a result,  
the stability threshold occurs for much lower (or even negative) scattering lengths~\cite{Lahaye2008}, departing significantly from $\epsilon_{dd}=1$. In that case the use of known results for homogeneous 3D BECs 
becomes unjustified. Due to this reason in all calculations below we restrict our analysis to aspect ratios $\lambda\leq 3$. 



\subsection{Gaussian Ansatz}

Although the droplets discussed below are in the TF regime, a qualitative, and to some extent quantitative, insight in the droplet physics may be gained from a simplified 
Gaussian ansatz~\cite{Yi2001}: 
\begin{equation}
 \psi(x,y,z)=\frac{1}{\pi^{\frac{3}{4}}\left(w_x w_y w_z\right)^{\frac{1}{2}}} \prod_{\eta=x,y,z} {\rm e}^{-\frac{\eta^2}{2w_\eta^2}+i\eta^2 \beta_{\eta}(t)}, \label{GaussianAnsatz}
\end{equation}
where the variational parameters are the widths $w_\eta$ in the $\eta=x,y,z$ direction, and $\beta_\eta$, which determines the phase curvature along $\eta$. 
The Lagrangian density reads
\begin{align}
 \mathcal{L}=&\frac{i \hbar}{2}\left(\psi \frac{\partial \psi^{*}({\bf r},t)}{\partial t}-\psi^{*}\frac{\partial \psi({\bf r},t)}{\partial t}\right)+\frac{\hbar^2}{2m}|\nabla \psi({\bf r},t)|^2 \nonumber \\ 
 &+V({\bf r})|\psi({\bf r},t)|^2+\frac{g}{2}|\psi({\bf r},t)|^4+\frac{2}{5}g_{\rm LHY}|\psi({\bf r},t)|^5\nonumber \\
 &+\frac{1}{2}\int d^3r' V_{\rm dd}({\bf r}-{\bf r}')|\psi({\bf r},t)|^2|\psi({\bf r}',t)|^2. 
 \label{LagrangianDensity}
\end{align}
We insert the ansatz~\eqref{GaussianAnsatz} into~\eqref{LagrangianDensity}, obtain the Lagrangian $L=\int d^3r \mathcal{L}$, and establish the corresponding 
Euler-Lagrange equations:
\begin{equation}
\beta_\eta=\frac{m}{2\hbar w_\eta} \frac{dw_\eta}{dt}
\end{equation}
and 
\begin{equation}
\frac{d^2 v_j}{d \tau^2}=-\frac{\partial}{\partial v_j}U(v_x,v_y,v_z). 
\end{equation}
In the latter equation, we have employed dimensionless units $\tau=\tilde{\omega}t $, $w_\eta=\tilde{l}v_\eta$, $\tilde{l}=\sqrt{\hbar/m \tilde{\omega}}$, with $\tilde\omega=(\prod\omega_\eta)^{1/3}$, 
and we have introduced the effective potential:
\begin{eqnarray}
 U&=& \frac{1}{2}\sum_{\eta}\left [ v_\eta^{-2}+\left ( \frac{\omega_\eta}{\tilde\omega} \right )^2v_\eta^2 \right ]+\frac{2}{3}\frac{PQ}{\left(\prod_\eta v_\eta \right)^{\frac{3}{2}}}\nonumber \\
 &+&  \frac{P}{\prod_\eta v_\eta} \left ( 1+\epsilon_{dd}F \left (\frac{w_z}{w_x},\frac{w_z}{w_y} \right )\right ),
 \label{Energy_vi}
\end{eqnarray}
where 
\begin{eqnarray}
&&F(\kappa_x,\kappa_y)=\frac{1}{4\pi}\int_0^\pi d\theta \sin\theta \int_0^{2\pi} d\phi \nonumber \\
&&\left [ \frac{3\cos^2\theta}{\left ( \kappa_x^2\cos^2\phi+\kappa_y^2\sin^2\phi \right) \sin^2\theta+\cos^2\theta}-1 \right ],
\end{eqnarray}
and the dimensionless constants $P=\sqrt{\frac{2}{\pi}}\frac{Na}{\tilde{l}}$ and 
 $Q=\frac{512\, F(\epsilon_{dd})}{25\sqrt{5}\pi^{\frac{5}{4}}}\sqrt{N\frac{a^3}{\tilde{l}^3}}$ characterize, respectively, the strength of the contact interaction and the LHY correction.
 
The equilibrium widths $v_{i0}$ are calculated by minimizing the potential $U$. In addition, 
the low-lying excitations around the equilibrium are determined by evaluating the Hessian matrix $M_{\eta\eta'}=\frac{\partial^2 U}{\partial v_\eta \partial v_{\eta'}}$ at the minimum. 
We consider below for simplicity a cylindrical trap, $\omega_x=\omega_y=\omega_z/\lambda$. In that case, the lowest excitation frequencies are:
\begin{eqnarray}
 \omega_1^2&=&M_{xx}-M_{xy} \label{eq:w1}\\
 \omega_{2,3}^2 &=& \frac{1}{2} \left ( M_{xx} + M_{xy} + M_{zz} \right) \nonumber \\
& \pm& \frac{1}{2} \sqrt{ (M_{xx}+M_{xy}-M_{zz})^2 + 8 M_{xz}^2 } \label{eq:w23}
\end{eqnarray}
The corresponding eigenvectors $(\eta_x,\eta_y,\eta_z)$ characterize the mode geometry. In particular, a mode with $\text{sign} (\eta_{x,y,z})=\pm$ has a 3D monopole character; 
$\text{sign}( \eta_{x,y})=\pm$ and $\text{sign}(\eta_z)=\mp$ a 3D quadrupolar character. The mode $(1\sqrt{2},-1/\sqrt{2},0)$ is a 2D quadrupolar mode on the $xy$ plane.

\section{Ground state}
\label{sec:Ground}

\subsection{Droplet versus mean-field solution}

We discuss in this section the ground-state properties of the dipolar BEC in the presence of LHY stabilization. 
We consider a trap with $\tilde{\omega}/2\pi=70\,$Hz, a similar value as that of recent Dy experiments~\cite{Kadau2015}. 
For a given aspect ratio $\lambda$ we obtain the ground state by imaginary time evolution~(ITE) of Eq.~\eqref{GPE}, using 
the split operator technique, treating the DDI using convolution theorem and fast-Fourier transformation, and cutting-off of the DDI to reduce spurious boundary 
effects~\cite{Ronen2006}. In absence of LHY stabilization, there is a critical $a_{cr}(\lambda)$, such that for $a<a_{cr}$ the dipolar BEC becomes unstable against collapse, i.e. no 
stable (or metastable) ground state exists. In contrast, due to the LHY term, Eq.~\eqref{GPE} presents for any value of $a$ and $\lambda$ a ground-state solution. This solution depends 
however crucially on $N$, $a$, and $\lambda$.



\begin{figure}[t]
 \begin{center}
 \includegraphics[width=\columnwidth]{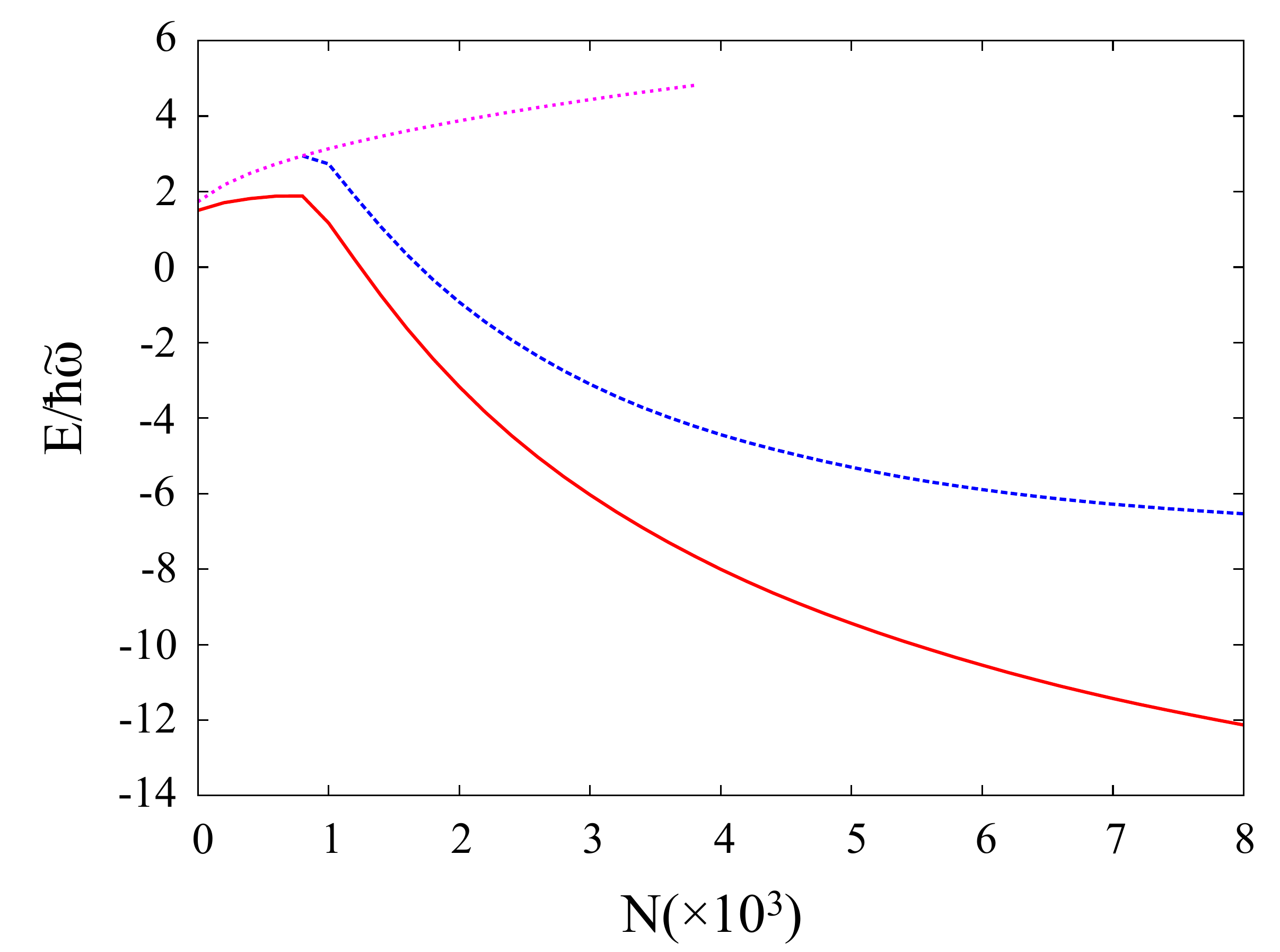}
\end{center}
\caption{(Color online) Energy per particle for a Dy BEC with $a=70a_B$ in a trap with $\tilde{\omega}/2\pi=70$Hz for $\lambda=1$~(solid), and for 
the droplet~(dashed) and mean-field~(dotted) solutions for $\lambda=3$.}
\label{fig:1}
\end{figure}




\begin{figure}[t]
 \begin{center}
 \includegraphics[width=\columnwidth]{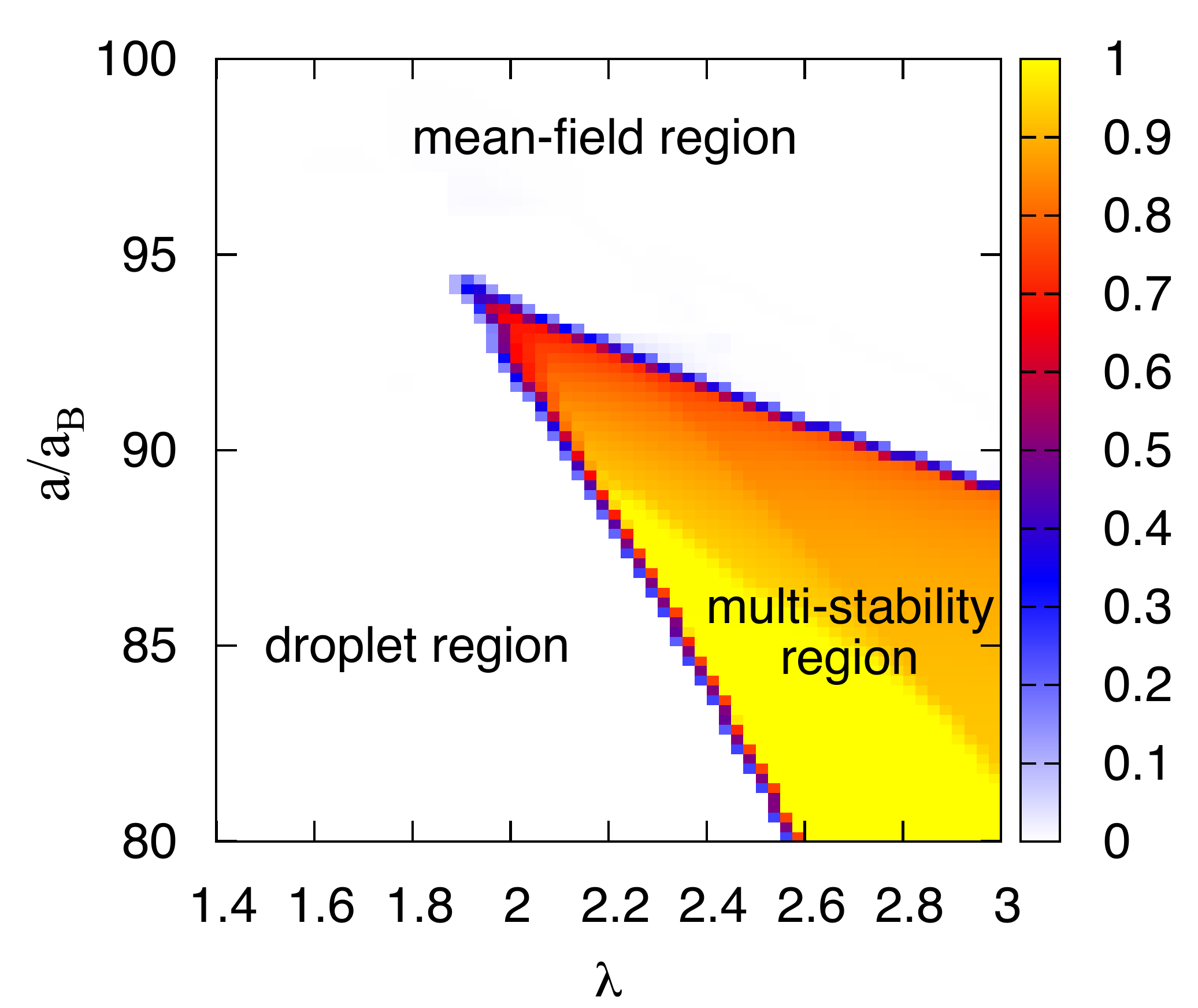}
\end{center}
\caption{(Color online) Ground-state phase diagram for $N=20000$ Dy atoms in a cylindrically harmonic trap with $\tilde{\omega}/2\pi=70\,$ Hz, as a function of the scattering length $a$ and the trap aspect ratio $\lambda$.
In the multi-stability region we depict the relative difference between the peak densities of the droplet and the mean-field solutions, $|n_P^{(D)}-n_P^{(M)}|/(n_P^{(D)}+n_P^{(M)})$.}
\label{fig:2}
\end{figure}


We depict in Fig.~\ref{fig:1}~(solid curve) the energy per particle as a function of $N$ for $a=70a_B$ and a spherical trap, $\lambda=1$. In absence of LHY stabilization, and similar to 
the case of non-dipolar gases with $a<0$, small condensates are stabilized by zero-point oscillations, rather than by the LHY term. This is still the case in Fig.~\ref{fig:1} for $N<1000$. In contrast, 
for $N>1000$ the LHY term is crucial to stabilize the cloud, the condensate becomes elongated along the dipole direction, and the BEC energy decreases to markedly negative values. 
Henceforth we call this elongated solution the {\em droplet solution}.
Whereas for $\lambda=1$ there is just one possible solution in the ITE of Eq.~\eqref{GPE},  the situation is clearly different for $\lambda=3$. 
For $N>1000$ the minimal energy solution~(dashed curve) is provided by an elongated solution of negative energy, similar to that found for $\lambda=1$. There is however a continuum of 
metastable states with different number of droplets with variable number of particles. More relevantly, there is a metastable state~(dotted curve) that connects smoothly with the solution for $N<1000$. 
This solution has a pancake geometry, being wider on the $xy$ plane than along $z$, as one would expect for a stable BEC in absence of LHY correction. 
The metastable pancake solution exists up to $N\simeq 4000$. 
In the following, and in order to discern it from the droplet solution, we call this solution the {\em mean-field solution}~(although the LHY term 
may play already a non-negligible role in its properties).


\subsection{Multi-stability}

The presence of metastable states marks a clear difference between cigar-like and pancake-like traps, which is best illustrated by the dependence of the BEC physics on $a$ and $\lambda$. 
Figure~\ref{fig:2} summarizes our results for $N=20000$ Dy atoms.  For large $a$, the condensate does not require LHY stabilization, being mean-field stable. Only the mean-field solution exists. 
For $\lambda<\lambda_{cr}$~($\lambda_{cr}\simeq 1.8$ in Fig.~\ref{fig:2}), there is just one ITE solution that smoothly connects for decreasing $a$ the mean-field and the droplet solution. As shown below, this 
smooth crossover results in a characteristic growth of the peak density. 

For $\lambda>\lambda_{cr}$ there is a region of $a$ values where {\em multi-stability} occurs, that 
separates the mean-field and droplet regions. Within the multistable region the color code in Fig.~\ref{fig:2}
depicts the relative difference $|n_P^{(D)}-n_P^{(M)}|/(n_P^{(D)}+n_P^{(M)})$, between the peak densities of the droplet solution, $n_P^{(D)}$, and the mean-field solution, $n_P{(M)}$. 
The lowest border of this region marks the end of the metastability of the mean-field solution.
The upper border marks a first order phase transition, characterized by a kink in the chemical potential of the ground-state solution, at which the droplet solution becomes the global energy minimum. 
We note at this point that three-body stabilization leads to a similar ground-state diagram with mean-field, droplet, and multi-stable regions~\cite{Blakie2016}.


\begin{figure}[t]
 \begin{center}
 \includegraphics[width=\columnwidth]{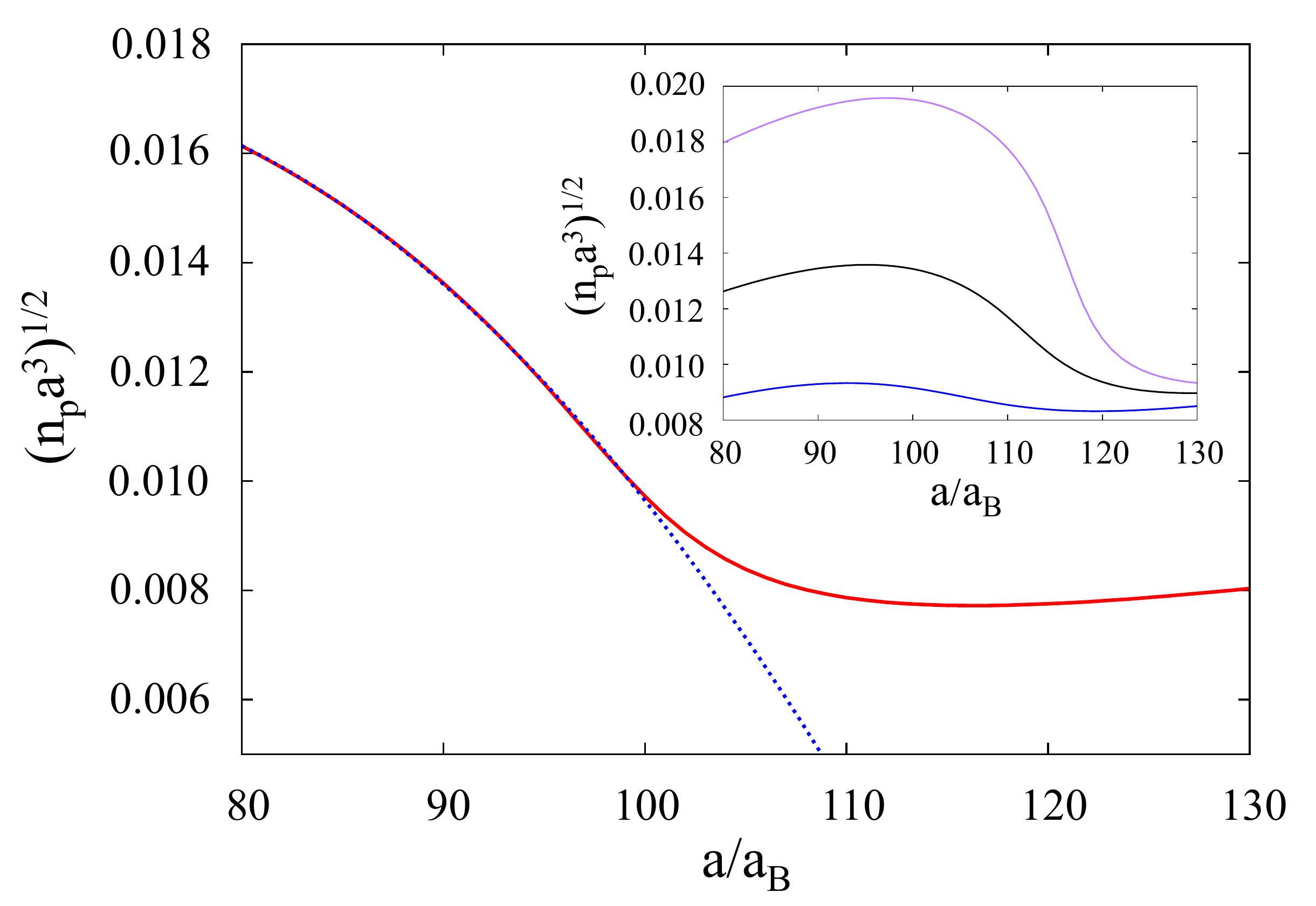}
\end{center}
\caption{(Color online) $(n_P a^3)^{1/2}$ for a Dy BEC of $N=20000$ in a spherical trap with $\tilde{\omega}/2\pi=70\,$ Hz as a function of the scattering length $a$. 
The curve follows in the droplet region the dependence $A\left(1-a/a_c\right)+B\left(1-a/a_c\right)^2$ 
with $A=0.082$, $B=-0.097$ and $a_c=116.48 a_B$~(dotted curve). In the inset we depict the results obtained for three-body stabilization with 
(from bottom to top) $g_3=2$, $1$, and $0.5 \times 10^{-38}m^6/s$.}
\label{fig:3}
\end{figure}


\subsection{Crossover from the mean-field to the droplet solution}

As mentioned above, for $\lambda<\lambda_{cr}$ there is a smooth crossover for decreasing $a$ between the mean-field and the droplet solution marked by a characteristic growth of the peak density. 
This growth may be employed to discern between LHY stabilization and stabilization based on strong conservative three-body forces~\cite{footnote-growth}. 
The latter would involve a term of the form $\frac{ \hbar g_3}{2}|\psi({\mathbf r,t})|^4$ in Eq.~\eqref{GPE} instead of the LHY term~\cite{Xi2015,Bisset2015}. 

Figure~\ref{fig:3} shows our results based on ITE of Eq.~\eqref{GPE} for the peak density as a function of $a$ for $N=20000$ Dy atoms in a spherical trap, $\lambda=1$. 
The peak density $n_P$ increases dramatically when $a$ decreases and the BEC enters the droplet region.
The dependence on $a$ of the peak density is markedly different for the LHY and three-body stabilization mechanisms. 
This difference is more evident when comparing the functional form of $(n_P a^3)^{1/2}$. 
As shown in Fig.~\ref{fig:3}, for the LHY stabilization $(n_P a^3)^{1/2}$ follows in the droplet region a characteristic dependence $A\left(1-a/a_c\right)+B\left(1-a/a_c\right)^2$, with $A$, $B$, and $a_c$ 
fitting parameters~\cite{footnote-dependence}. This functional form fits well the peak density for all $\lambda$ and particle numbers within the droplet regime. We note as well that the qualitative behavior of the 
peak-density scaling is also in good agreement with the results obtained from the Gaussian ansatz discussed above.
This dependence is clearly lost in the case of three-body stabilization, irrespective of the value of $g_3$~(inset of Fig.~\ref{fig:3}). 
Hence the analysis of $(n_Pa^3)^{1/2}$ as a function of $a$ in the crossover regime~($\lambda<\lambda_{cr}$) 
provides a clear way to discern between the two stabilization mechanisms.



\begin{figure}[t]
 \begin{center}
 \includegraphics[width=\columnwidth]{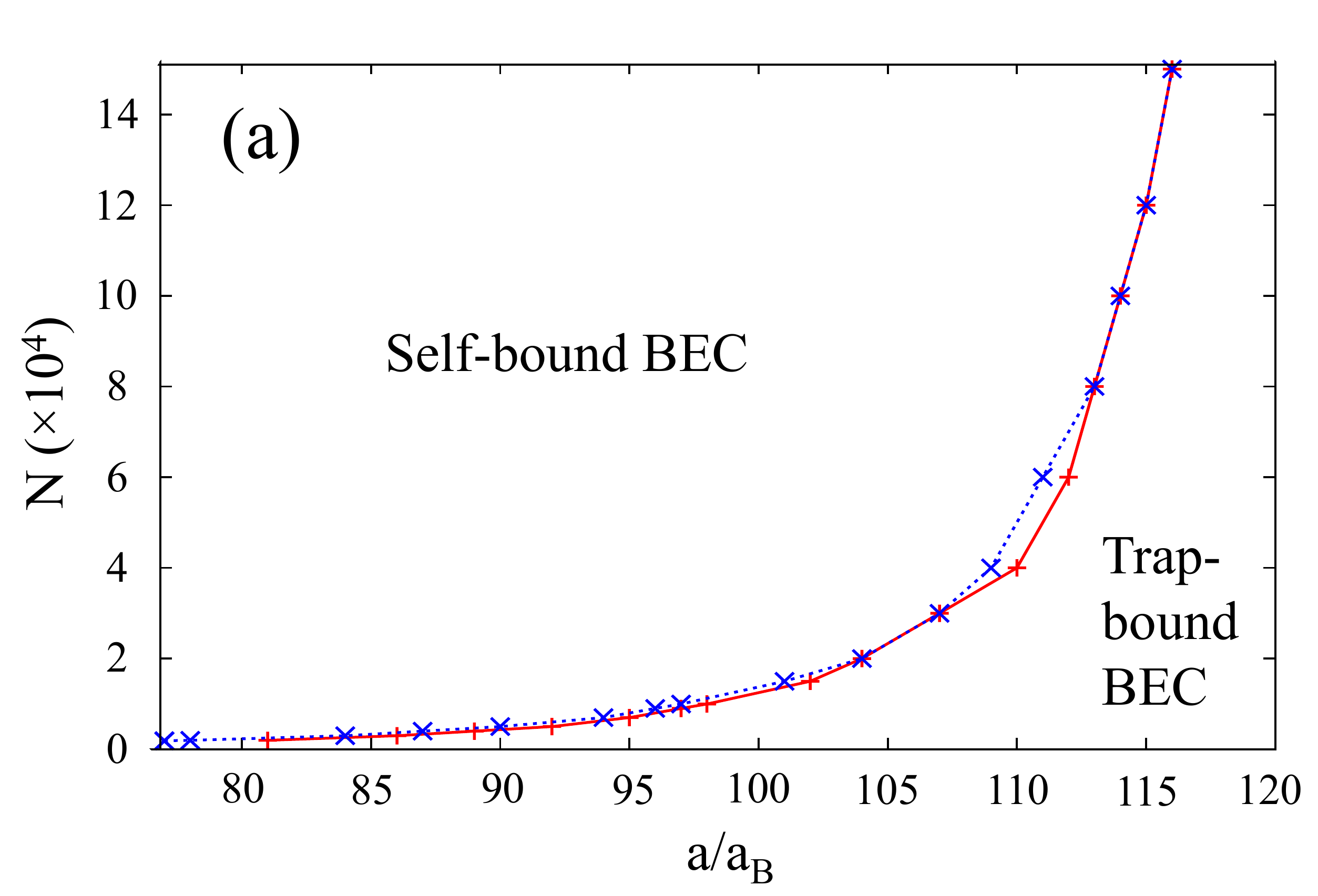}
 \includegraphics[width=\columnwidth]{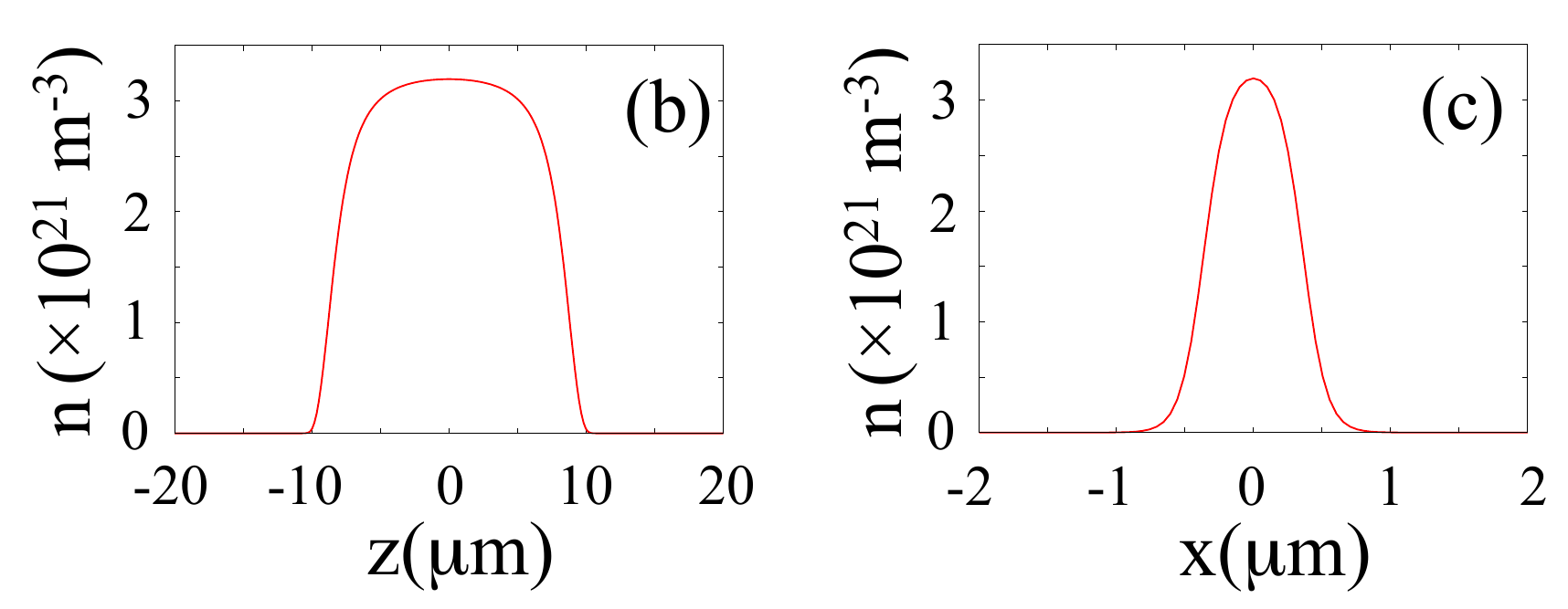}
\end{center}
\caption{(Color online) (a) Self-bound and trapped-bound regions for a Dy BEC as a function of the number of particles $N$ and the scattering length $a$. The figure 
shows the boundary as calculated from ITE of Eq.~\eqref{GPE}~($\times$) and from the Gaussian variational Ansatz~($+$).  
The lower figures show the density profile along $z$ for $x=y=0$~(b) and along $x$ for $y=z=0$~(c), for a Dy BEC with $N=20000$ and $a=80a_B$, well within the SB regime.}
\label{fig:4}
\end{figure}




\begin{figure}[t]
\includegraphics[width=1.0\columnwidth]{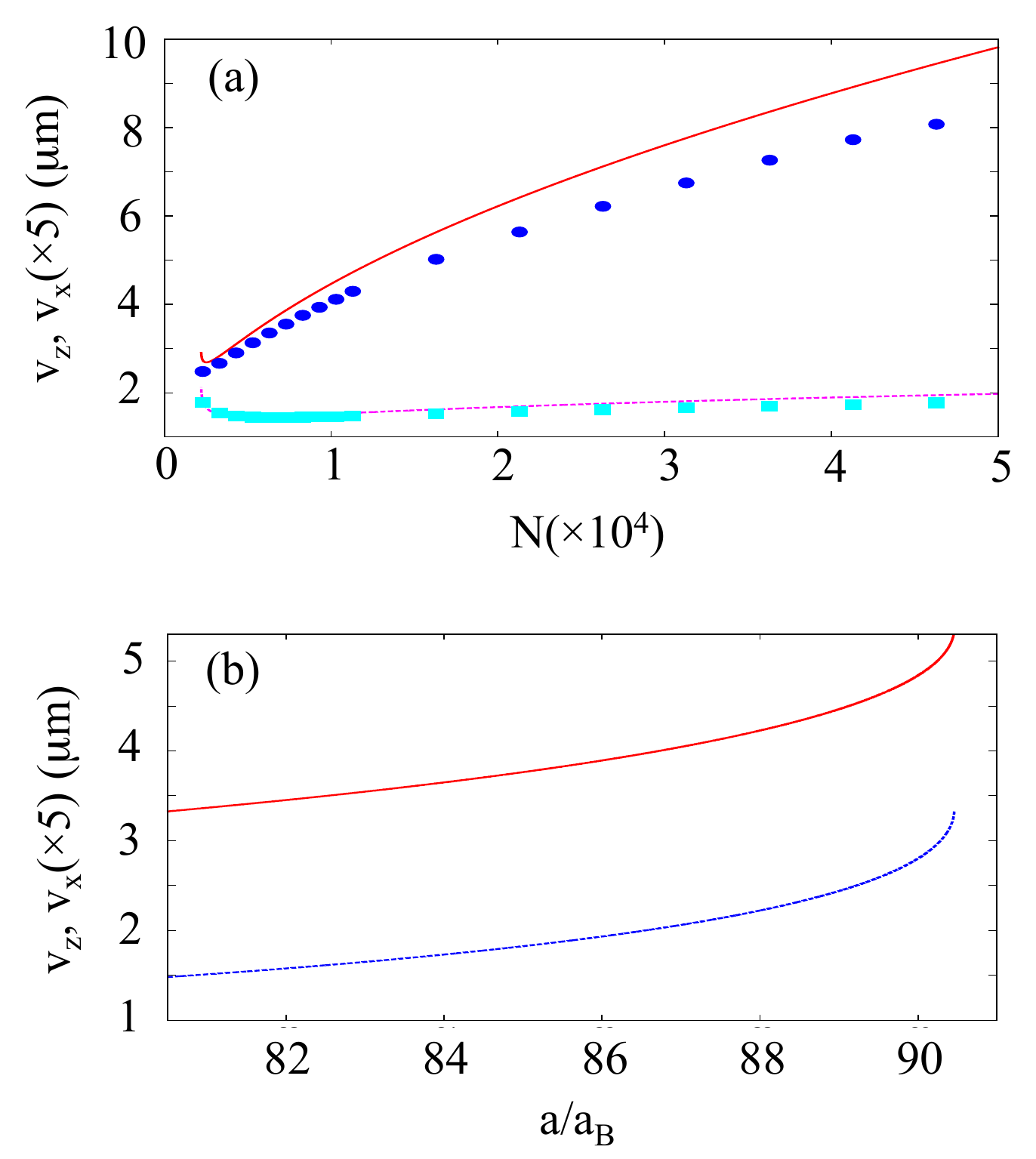}
\vspace*{-0.4cm}
\caption{(Color online) (a) $v_z$~(solid) and $v_x$~(dashed) for a SB solution with $a=80 a_B$ as a function of $N$; 
(b) Same but for a fixed $N=5000$ as a function of $a$. The curves depict the results obtained from the variational Gaussian Ansatz. 
In Fig. (a) we show as well the results obtained from the direct simulation of Eq.~\eqref{GPE} for $\sqrt{2 \langle z^2\rangle }$~(circles) and $\sqrt{2 \langle x^2\rangle }$~(squares)}
\label{fig:5}
\end{figure}



\section{Self-bound condensates}
\label{sec:SB}


The interplay of LHY and mean-field terms allows for 3D self-bound~(SB) condensates for a sufficiently small $a$. 
Figure~\ref{fig:4}(a) shows the boundary between SB and trap-bound solutions. The curve marked by $\times$ symbols is obtained as the point in which 
ITE of Eq.~\eqref{GPE} results in an unbound solution. The curve marked with $+$ symbols is evaluated from the simplified Gaussian ansatz as the point at which the minimum of the 
effective potential $U$ disappears. Both curves are in excellent agreement. Note that for larger number of particles, the boundary is basically vertical, marking a critical scattering length 
$a_{SB}$, such that for $a>a_{SB}$ no self-bound solution is possible~(for Dy, $a_{SB}\simeq 120 a_B$).

It is important to stress, that within the SB regime, and for a sufficiently large number of particles, the SB droplets are approximately in the TF regime, but obviously they do not present
the typical inverted-paraboloid density profile due to the modified equation of state and the absence of harmonic confinement~(Figs.~\ref{fig:4}(b) and (c)). This constitutes a clear difference between 
these SB BECs and bright BEC solitons. The latter are also SB solutions, which however just exist in 1D~(or in 2D for dipolar BECs~\cite{Pedri2005}), resulting from the compensation of 
quantum pressure and attractive mean-field interactions. Hence by definition they cannot exist in the TF regime. 
Here, in contrast, the droplet remains SB in 3D by the compensation of LHY and mean-field terms, and hence TF SB droplets are allowed, being in fact the general case. 

Figure~\ref{fig:4} shows that there are two non-equivalent ways of entering the SB regime, either increasing $N$ or decreasing $a$. 
Figure~\ref{fig:5} depicts the widths $v_{x,z}^{SB}$ of the SB BECs. 
Decreasing $a$ at constant $N$ leads to a smaller $v_{x,z}^{SB}$~(Fig.~\ref{fig:5}(b)), although both $\kappa$ and $n_P$ increase significantly. 
In contrast, increasing $N$ for constant $a$ results in a rapid increase of $v_z^{SB}$~(Fig.~\ref{fig:5}(a)), whereas $v_x^{SB}$ remains almost constant, 
and $n_P$ increases~(note that the variational results and those obtained from the direct simulation of Eq.~\eqref{GPE} are in good agreement, despite 
the clearly non-Gaussian nature of the BEC deep inside the SB regime). This dependence is relevant for the convergence of the trapped BEC to the SB solution, 
since as discussed in the following, the convergence of the trapped solution to the SB solution is eased if $v_z^{SB}$ is small. 
The realization of three-dimensional SB BEC is hence significantly simpler when working at small $a$ and low $N$.

\begin{figure}[t]
\includegraphics[width=1.0\linewidth]{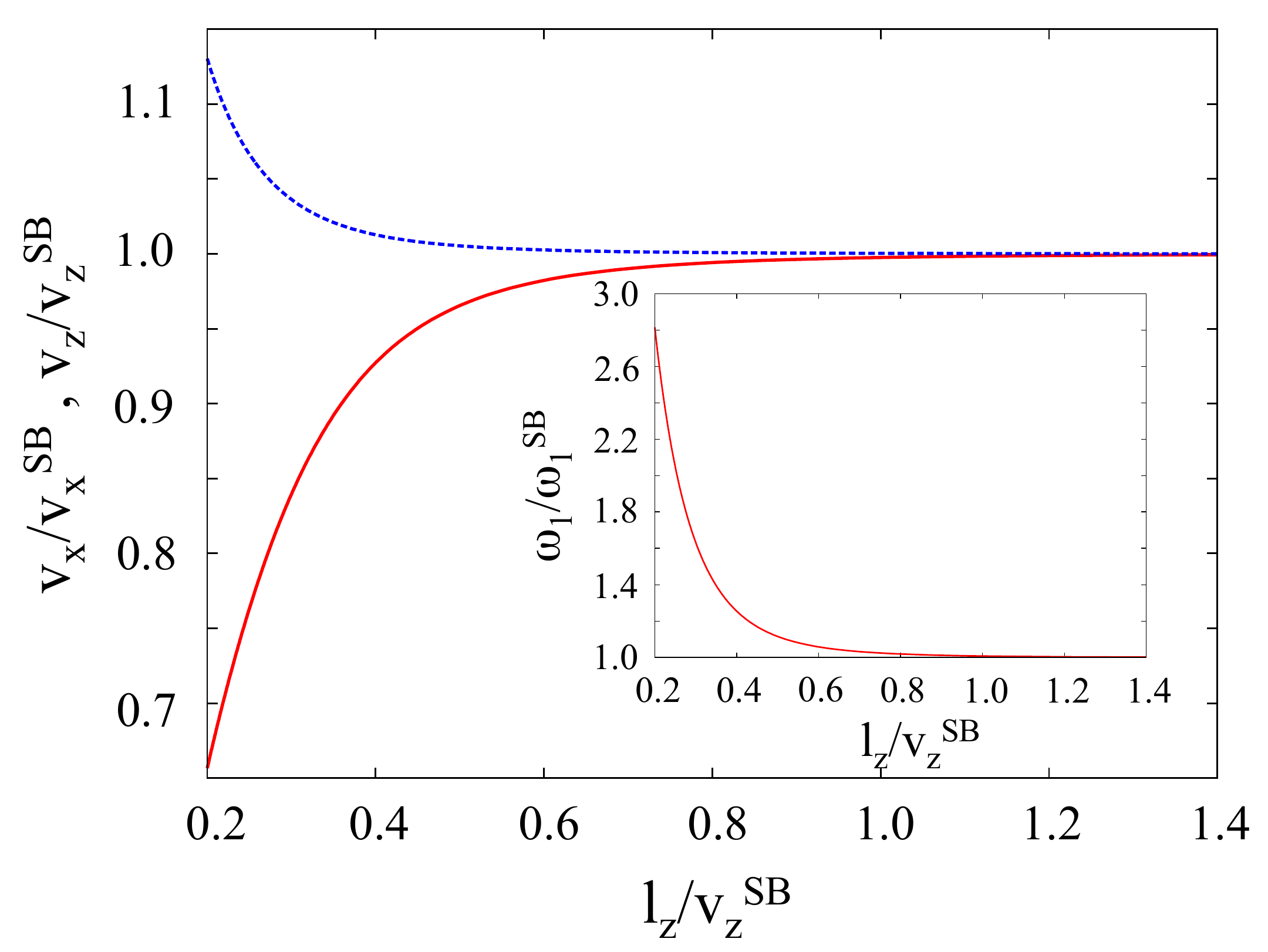}
\vspace*{-0.4cm}
\caption{(Color online) Widths $v_z$~(solid) and $v_x$~(dashed)~(obtained using the variational Gaussian ansatz) for a Dy condensate in a spherical trap~($\lambda=1$) 
with $N=5000$ atoms, and $a=80a_B$. The results are normalized to the corresponding widths $v_z^{SB}$ and $v_x^{SB}$ of the SB solution for those parameters, 
and plotted as a function of the ratio $l_z/v_z^{SB}$, with $l_z$ the oscillator length of the $z$ confinement. Inset: lowest excitation mode for the same parameters. 
For these parameters $v_x^{SB}=0.3\mu$m and $v_z^{SB}=3.3\mu$m.}
\label{fig:6}
\end{figure}

The trap may significantly alter the properties of the SB solution. Due to the marked elongation of the SB droplets 
along $z$, the $z$ confinement is particularly crucial. If the corresponding oscillator length $l_z<v_z^{SB}$, the BEC may depart significantly from the SB solution. 
This is true not only for the $z$-width of the cloud, but also for the radial one, despite the fact that the condensate is much narrower radially. 
Figure~\ref{fig:6} shows our Gaussian Ansatz results for $v_z/v_z^{SB}$ and $v_x/v_x^{SB}$ for $\lambda=1$, $N=5000$ Dy atoms, and $a=80a_B$, 
as a function of  the ratio $l_z/v_z^{SB}$. Convergence demands $l_z/v_z^{SB}>1$, which for this case would demand a rather low $\omega_z/2\pi<5.6$Hz. 
For typical experimental values of $\tilde \omega/2\pi=70$Hz, $v_z/v_z^{SB}\simeq 0.8$, $v_x/v_x^{SB}\simeq 1.05$, and $n_P/n_P^{SB}\simeq 1.12$, and hence 
the deviation from the SB solution is relatively small. As discussed above, the realization of the SB solution is much more involved for larger $N$ and $a$. 
For example, for $N=20000$ and $a=100a_B$, for a spherical trap with $\tilde \omega/2\pi=70$Hz, 
$v_z/v_z^{SB}\simeq 0.6$, $v_x/v_x^{SB}\simeq 1.2$, and $n_P/n_P^{SB}\simeq 1.06$.
Typical experiments would hence produce BECs that albeit stabilized by the LHY term may be well away from the 3D SB regime. 
As a result, abruptly switching the trap is typically not expected to result in a complete cancellation of the time-of-flight expansion, as one would expect from the 3D SB 
character~\cite{footnote-TOF-Saito}.



\begin{figure}[t]
 \begin{center}
 \includegraphics[width=\columnwidth]{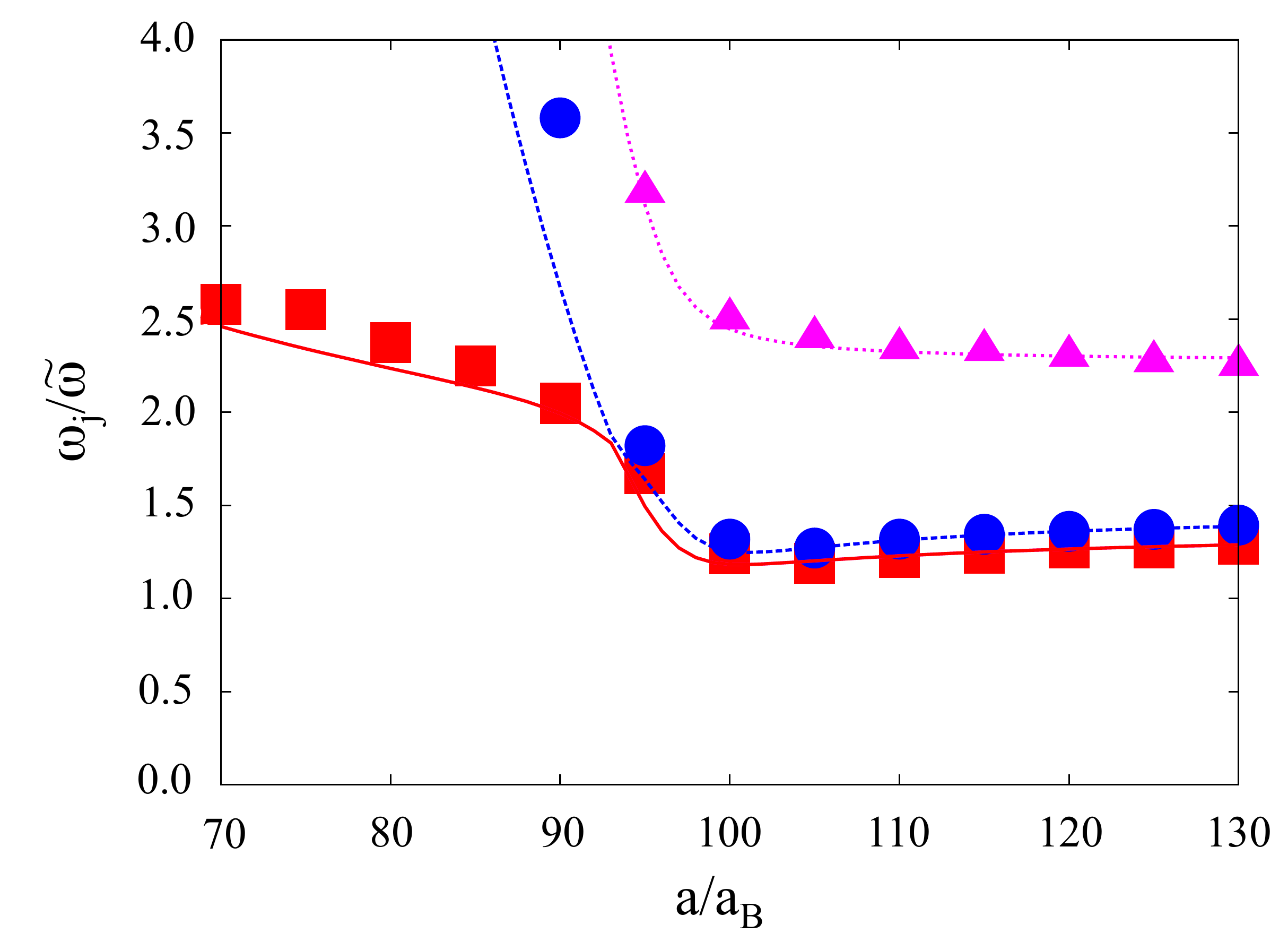}
\end{center}
\caption{Lowest excitation frequencies of a spherically trapped Dy BEC with $N=20000$ atoms and $\tilde\omega/2\pi=70$Hz. 
The numerical results for the three lowest modes are marked, respectively, by squares, circles, and triangles.
These results are obtained by monitoring the condensate after a slight trapping quench ($\epsilon=0.01$,  $\omega_{x,y,z}\rightarrow(1+\epsilon)\omega_{x,y,z}$). 
The corresponding frequencies evaluated from the Gaussian ansatz are depicted by, respectively, solid, dashed, and dotted lines.}
\label{fig:7}
\end{figure}


\section{Excitations}
\label{sec:Excitations}

Once established the ground state properties of the dipolar BEC in presence of the LHY stabilization, we focus 
on the lowest eigenmodes of the condensates. We evaluate the lowest-lying excitations 
from the condensate response to an abrupt small change of the trap frequencies $\omega_{x,y,z}=(1+\epsilon)\omega_{x,y,z}$, with $\epsilon=0.01$.
The subsequent dynamics is evaluated by real-time evolution of Eq.~\eqref{GPE}. We monitor in particular the variances of the BEC along each of the three spatial directions. 
The Fourier analysis of these variances reveals the underlying frequencies depicted in Fig.~\ref{fig:7} for the case of a spherically trapped Dy BEC with $N=20000$ atoms. 
The curves in Fig.~\ref{fig:7} depict the results obtained from Eqs.~\eqref{eq:w1} and~\eqref{eq:w23}, which are in very good agreement with the numerical results.

Deep in the mean-field regime~($a=130a_B$ in Fig.~\ref{fig:7}) the lowest excitation is given by a  radial quadrupole mode, $(1/\sqrt{2},-1/\sqrt{2},0)$, where we employ the eigenvector 
notation introduced  in Sec.~\ref{sec:Model}. The energetically second lowest mode is a 3D quadrupolar mode, $(-0.53,-0.53,0.66)$,  and the third one is a monopole-like mode, $(0.5,0.5,1/\sqrt{2})$. 
The character of these modes changes when the BEC crossovers into the droplet regime, $a<90 a_B$ in Fig.~\ref{fig:7}.
Due to the marked elongation of the droplet along the dipole direction, axial~(along $z$) and radial~(on the $xy$ plane) modes approximately decouple.
The 3D quadrupole-like mode becomes the lowest lying one, but it becomes almost completely an axial mode along $z$, $(-0.05,-0.05,0.99)$. 
The radial quadrupole mode does not change its character, whereas the 3D monopole-like mode becomes approximately a 2D monopole mode~$(0.7,0.7,0.07)$. 
Due to the large aspect ratio of the droplet, the quasi-radial modes become much more energetic than the quasi-axial mode. 
As a consequence, in the droplet regime, a slight quench of the trap frequencies just excites the lowest mode.

\begin{figure}[t]
\includegraphics[width=1.0\linewidth]{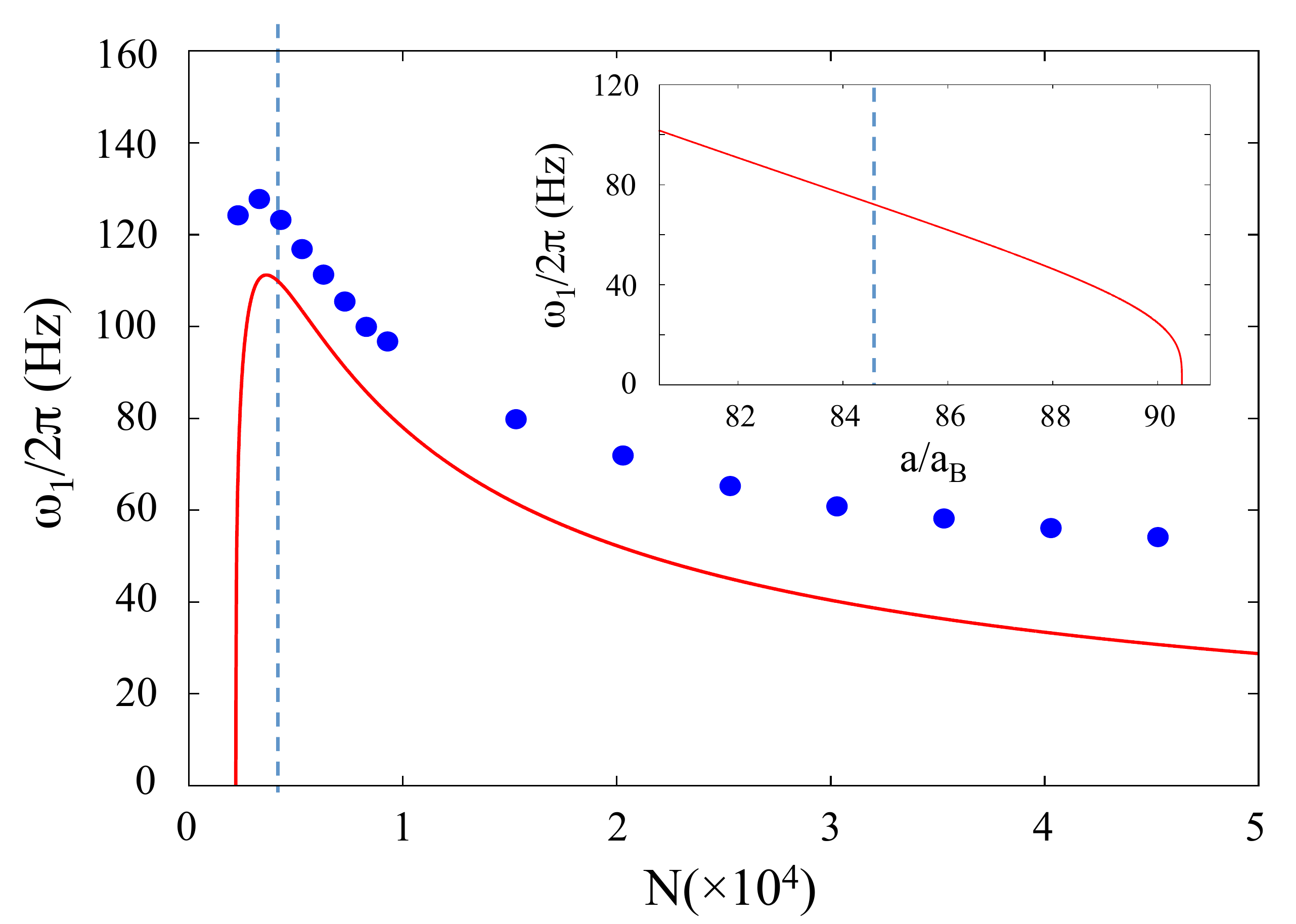}
\vspace*{-0.4cm}
\caption{(Color online) Lowest eigenenergy of the SB solution for a Dy BEC as a function of $N$ for $a=80a_B$; inset: same as a function of 
$a$ for $N=5000$. The dashed lines mark the point at which the lowest mode changes from a quadrupole to a monopole character before becoming unstable.
The curves have been obtained using the variational Gaussian ansatz, whereas the circles are obtained directly from the numerical simulation of Eq.~\eqref{GPE}.}
\label{fig:8}
\end{figure}

Due to similar reasons, in a SB BEC the lowest mode remains quasi-axial. Deep in the SB regime, this mode retains a slight 3D quadrupole character, as for the case of 
trapped BECs. However, close to the instability the mode becomes 3D monopole-like. The softening of this mode marks the unbinding of the dipolar BEC. 
Figure~\ref{fig:8} shows that whereas the lowest eigen-energy of the SB solution grows monotonously with decreasing $a$~(inset of Fig.~\ref{fig:8}), 
it does present a maximal value as a function of $N$, decreasing at large $N$. At this maximum the mode changes, for decreasing $N$, from quadrupole to monopole character.
Note that the Gaussian ansatz describes well the qualitative dependence of the excitation energy, although 
the quantitative value may significantly differ due to the clearly non-Gaussian nature of the BEC deep inside the SB regime~(see Figs.~\ref{fig:4}(b) and (c)).

Finally, let us note that the frequency of the lowest-lying mode of the trapped droplet departs significantly from that of the SB solution if $l_z/v_z^{SB}<1$~(inset of Fig.~\ref{fig:6}). 
For $N=5000$ Dy atoms in a spherical trap with $\tilde\omega/2\pi=70$ and $a=80a_B$, $\omega_1/\omega_1^{SB}\simeq 1.85$~(we recall that for this case, $v_z$, $v_x$ and $n_P$ presented a relatively 
small departure from the SB values).



\section{Conclusions}
\label{sec:Conclusions}

In this paper we have analyzed in detail the ground-state properties and lowest-lying excitations of a dipolar BEC in the presence of LHY stabilization. 
For a trap aspect ratio $\lambda>\lambda_{cr}$ the BEC presents three marked regions, a mean-field region, in which the LHY term is perturbative, 
a droplet region, and an intermediate multi-stable region. The mean-field to multi-stable boundary is characterized by a first order transition, at which the 
droplet solution becomes the global ground state. For $\lambda<\lambda_{cr}$ there is a crossover between the mean-field solution and the droplet one,  
marked by a characteristic functional dependence with $a$ of the peak density and of the lowest-lying excitation. Although we have focused in this paper 
on the particular case of dysprosium, similar results characterizes other strongly dipolar gases, in particular recent erbium experiments~\cite{Chomaz2016}. 

A major consequence of the quantum stabilization is the possibility to create three-dimensionally self-bound condensates, which would be hence  
characterized by a vanishing time-of-flight expansion velocity.  We have shown however 
that, due to the elongation of the BEC along the dipole direction, the convergence of the trapped solution to the self-bound one demands under typical conditions a rather 
weak confinement. If the confinement is not weak enough, the properties of the trapped BEC may significantly differ from the self-bound case, especially in what concerns 
the lowest-lying mode. As a result an abrupt switch off of the trap in time-of-flight experiments would create rather an excited solution, resulting in a finite time-of-flight expansion~\cite{Chomaz2016}. 
The observability of the 3D self-bound solution as a non-expanding condensate in time-of-flight experiments would hence demand BECs with small $N$ and $a$, and possibly 
a quasi-adiabatic opening of the trap prior to the TOF measurements. 

{\em Note added.--} In the process of finishing the writing of this paper we became aware of a recent preprint~\cite{Bisset2016}, where the 
ground-state phase diagram of a dipolar BEC with LHY stabilization is discussed, with similar conclusions concerning the three 
regions (mean-field, droplet, and multi-stable) discussed in Sec.~\ref{sec:Ground}. \\

\begin{acknowledgments}
We thank discussions with I. Ferrier-Barbut, T. Pfau, L. Chomaz, M. Mark, and F. Ferlaino.
We acknowledge support by the DFG Research Training Group 1729.
\end{acknowledgments}



\bibliographystyle{prsty}

\end{document}